\documentclass[aip,amsmath,amssymb,floatfix,citeautoscript,reprint]{revtex4-2}
\usepackage{cancel}
\usepackage{amsmath}
\usepackage{graphicx}
\usepackage{amssymb}
\usepackage{amsthm}
\usepackage{bm}
\usepackage{dcolumn}
\usepackage{braket}
\usepackage{longtable}
\usepackage{siunitx}

\usepackage{ragged2e}
\usepackage{txfonts}

\usepackage{color}
\usepackage[usenames,dvipsnames]{xcolor}
\definecolor{myblue}{rgb}{0,0,1}
\usepackage[breaklinks=true,colorlinks=true,linkcolor=myblue,urlcolor=myblue,citecolor=myblue]{hyperref}

\begin{document}

\title{Regional Embedding Enables High-Level Quantum Chemistry for Surface Science}

\author{Bryan T. G. Lau}
\affiliation{Center for Computational Quantum Physics, Flatiron Institute, New York, New York 10010 USA}
\author{Gerald Knizia}
\email{knizia@psu.edu}
\affiliation{Department of Chemistry, Pennsylvania State University, University Park, Pennsylvania 16802 USA}
\author{Timothy C. Berkelbach}
\email{tim.berkelbach@gmail.com}
\affiliation{Center for Computational Quantum Physics, Flatiron Institute, New York, New York 10010 USA}
\affiliation{Department of Chemistry, Columbia University, New York, New York 10027 USA}

\begin{abstract}

Compared to common density functionals, ab initio wave function methods can
provide greater reliability and accuracy, which could prove useful when modeling
adsorbates or defects of otherwise periodic systems.  However, the breaking of
translational symmetry necessitates large supercells that are often prohibitive
for correlated wave function methods.  As an alternative, we introduce the
regional embedding approach, which enables correlated wave function treatments
of \emph{only} a target fragment of interest through small, fragment-localized
orbital spaces constructed using a simple overlap criterion.  Applications to
the adsorption of water on lithium hydride, hexagonal boron nitride, and
graphene substrates show that regional embedding combined with focal point
corrections can provide converged CCSD(T) (coupled cluster) adsorption energies
with very small fragment sizes.

\end{abstract}

\maketitle

In electronic structure theory, ab initio wave function methods provide a
hierarchy of approximations that allow for systematic improvability--a feature
missing in density functional theory (DFT).  While ab initio methods were
originally limited to molecules, advances in the past decade have enabled
approaches like many-body perturbation theory or coupled-cluster theory to be
applied to atomistic periodic
systems.~\cite{Marsman2009,C2CP24020C,Booth2013,Yang2014,McClain2017,Gruber2018,Zhang2019,Dovesi2020}
However, these methods are expensive and exploitation of translational symmetry
is usually necessary to reduce their computational cost to manageable levels.
Unfortunately, this is impossible in several applications scenarios of
tremendous practical importance.  For example, in models of crystal defects
(e.g.~for doping in semiconductor devices), or surface adsorbates (e.g.~as
involved in most core processes of heterogeneous catalysis), translational
symmetry is broken.  In practice, the non-periodic system is then described with
supercells, which must be large in order to eliminate artifacts caused by the
spurious interactions of the defect or adsorbate with its periodic images in
other supercells.  This is particularly problematic for ab initio methods, which
are already expensive \emph{with} translational symmetry.

However, when translational symmetry is broken by spatially-localized
perturbations, the high cost of conventional treatments by large supercells is
often artifical and can be avoided.  To substantiate this claim, we here
demonstrate that ab initio molecular adsorption energies can be obtained at
significantly reduced cost, compared to the conventional supercell treatment, by
a regional embedding technique.  In general, the accurate prediction of
noncovalent adsorption energies is challenging because they arise as small
energy differences between large total energies~\cite{MAURER201672}.  However,
as opposed to total energies, adsorption energies are natural targets for
regional embedding approaches, because they focus the computational effort on
the important region of change.  Like previous periodic embedding
approaches~\cite{Govind1999,Chulhai2018,Virgus2014,Eskridge2019}, we only
correlate orbitals that are localized to a region near the adsorbate and can
converge to the supercell result by increasing the size of the region. This
technique allows us to perform expensive post-Hartree-Fock (HF) calculations on
\emph{only} the chemically important piece of the system, while retaining large
supercells at the mean-field level to eliminate finite-size and finite-coverage
effects.

Unlike most previous embedding approaches, we localize the orbitals based upon
their overlap with the fragment of interest in Hilbert space, rather than
invoking real-space densities or localization functionals.  The approach builds
on quantum information concepts and, in particular, density matrix embedding
theory (DMET)~\cite{Knizia2013,Wouters2016,doi:10.1021/acs.jctc.9b00933}. DMET
gives a precise prescription for the factorization of a system's wavefunction
into active and inactive components, which we leverage to define a regional
embedding procedure.  For the orbitals occupied in the HF determinant, we define
an operator that projects onto a minimal atomic orbital (AO) basis of the atoms
in the fragment $A$,
\begin{equation}
\hat{P}^{\mathrm{occ}} = \sum_{\rho,\tau\in A} 
    |\rho\rangle [\mathbf{S}^{-1}]_{\rho\tau} \langle\tau|
\end{equation}
where $\rho,\tau$ are the minimal AO basis functions on the fragment and
$\mathbf{S}$ is the overlap matrix in the minimal AO basis; let $|a|$ indicate
the total number of minimal AO basis functions on the fragment. 
The eigenvectors of
$\hat{P}^{\mathrm{occ}}$ in the
basis of canonical occupied orbitals,
$P_{ij}^{\mathrm{occ}} 
= \langle\varphi_i|\hat{P}^{\mathrm{occ}}|\varphi_j\rangle$, 
define a rotated set of occupied orbitals whose eigenvalues quantify their
overlap with the fragment space.  As a unitary transformation of the occupied
orbitals, this procedure does not change the HF wavefunction, but merely
separates the occupied orbitals into a set with nonzero overlap on the fragment
(at most $|a|$) and a set with zero overlap on the fragment.  If we were to use
the same projection operator to rotate the canonical virtual orbitals, this
procedure would generate the atomic valence active space~\cite{Sayfutyarova2017}
of the fragment atoms.  As explained by DMET, the combined active space would
have at most $2|a|$ orbitals, which is the maximum number required to represent
the mean-field hybridization between the fragment minimal AOs and the rest of
the system.  However, this small active space lacks the basis functions
necessary for a description of dynamical correlation, which is critical for
accurate adsorption energies.  Therefore, we modify the procedure for the
virtual space. We now let the operator
$\hat{P}^{\mathrm{vir}}$ project onto the \textit{computational} AO basis
functions of the atoms in the fragment, and denote the number of these basis
functions as $|A|$.  Diagonalization in the basis of the canonical virtual
orbitals, $P_{ab}^{\mathrm{vir}} =
\langle\varphi_a|\hat{P}^{\mathrm{vir}}|\varphi_b\rangle$, then defines
a unitary transformations separating the virtuals into those with nonzero
overlap on the fragment (at most $|A|$) and those with zero overlap on the
fragment.  As discussed in Refs.~\onlinecite{Zheng2016,Sayfutyarova2017}, the
combined occupied and virtual orbital spaces with fragment overlap are closely
related to the combined impurity and bath orbital spaces in conventional DMET.

We propose to correlate only those orbitals that have significant overlap with
the fragment, defined by a threshold on the eigenvalues of the projection
operators. The rotated orbitals with insufficient overlap on the fragment space
are simply frozen in the post-HF calculations, making the insertion of regional
embedding into existing quantum chemistry workflows straightforward.  Compared
to other localization procedures, the one proposed here does not require
minimization of a cost function (as in Foster-Boys~\cite{Foster1960},
Edmiston-Ruedenberg~\cite{Edmiston1963}, Pipek-Mezey~\cite{Pipek1989}, or
higher-order moment-based~\cite{Hoyvik2012,Hoyvik2014} localization) and does
not result in nonorthogonal orbitals (as in projected atomic
orbitals~\cite{PULAY1983151,doi:10.1146/annurev.pc.44.100193.001241,doi:10.1063/1.471289,doi:10.1063/1.1330207,Hoyvik2014}).
We note that two recent works have explored similar overlap-based approaches to
embedding, focusing on the occupied orbitals only~\cite{Claudino2019} and on
multireference problems~\cite{He2020}.  In future work we will compare the
spatial extent and convergence properties of the local orbitals of these various
approaches more thoroughly.

While any high-level quantum chemistry method may be used in the fragment
calculation, here we focus on second-order M\o ller-Plesset perturbation theory
(MP2) and coupled-cluster theory with single and double excitations (CCSD) and
perturbative triples (CCSD(T)). These methods are valuable in this problem area
for their nonempirical treatment of dispersion interactions.  Furthermore,
CCSD(T) is a powerful method in molecular thermochemistry: in small and
electronically benign molecules, energy differences computed at its basis set
limit typically deviate by $\leq$2 kJ/mol from the exact result, far
outperforming all known DFT methods~\cite{karton2006w4}.  The extent to which
this accuracy persists for the thermochemistry of solids and surfaces remains to
be seen.

To test regional embedding, we study the adsorption of a water molecule on the
(001) surface of lithium hydride (LiH)\cite{Booth2016,Tsatsoulis2017}, hexagonal
boron nitride (hBN)\cite{Al-Hamdani2017,Gruber2018}, and
graphene\cite{Voloshina2011,C000988A,Brandenburg2019,Jordan2019}.  These systems
have been used to demonstrate the application of ab initio methods to periodic
systems, and reference results obtained by extrapolation to the thermodynamic
and complete basis set limits are available for a variety of DFT, wavefunction,
and quantum Monte Carlo methods. We use literature values for the orientation
and location of the water molecule; for water on graphene, we consider the
so-called ``0-leg'' configuration~\cite{Brandenburg2019}.  The calculations are
performed with slab geometries using a single layer of graphene and hBN and two
layers of the LiH $(001)$ plane along with \SI{20}{\angstrom} of vacuum in the
direction perpendicular to the surface.

\begin{figure}
\includegraphics[scale=0.85]{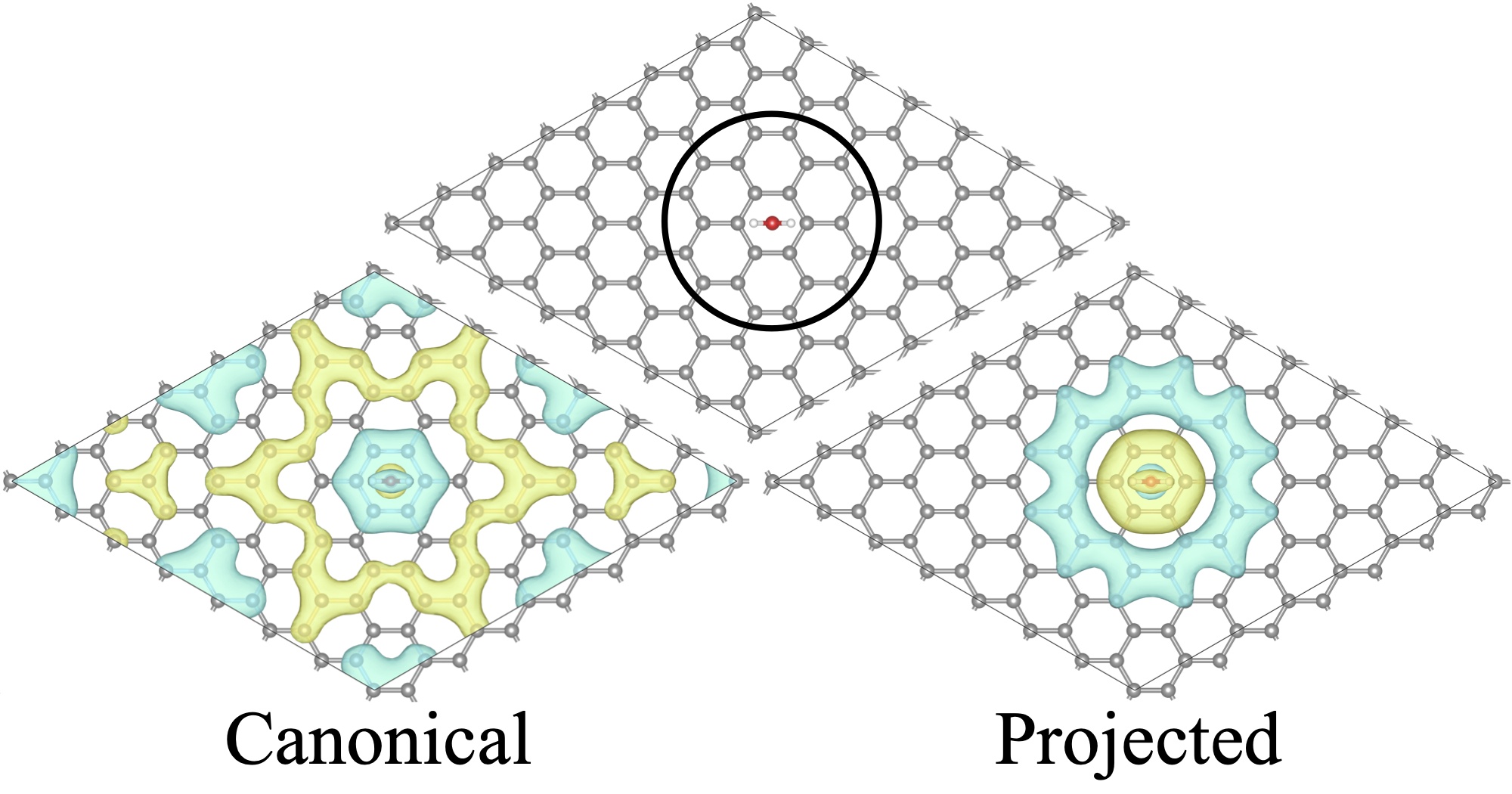}
\caption{An illustration of the embedding scheme used in this paper, for water
on a $7\times 7$ graphene supercell, with an example of randomly selected
canonical and projected occupied orbitals. Carbon atoms from graphene are
selected to be in the fragment based upon the radial distance from the oxygen
atom in the water molecule.}
\label{fig:1}
\end{figure}

Figure~\ref{fig:1} illustrates the method of selecting atoms in the fragment,
using the example of water on graphene.  The atoms of the water molecule are
always included in the fragment and we add atoms from the substrate based upon
their radial distance from the water molecule.  The adsorption energy is found
from three separate calculations,
\begin{equation}
E_\mathrm{ads}
    = E_\mathrm{H_2 O+substrate} - E_\mathrm{H_2 O} - E_\mathrm{substrate}
\end{equation}
where the embedding procedure is applied to each calculation. We apply the
counterpoise correction for basis set superposition error by performing each
calculation in the full basis set of the water and
substrate~\cite{doi:10.1080/00268977000101561}, using the same eigenvalue cutoff
in each.

All calculations are performed with periodic boundary conditions as implemented
in PySCF~\cite{doi:10.1002/wcms.1340,Sun_2020}, using Gaussian density
fitting~\cite{Sun2017} of periodic integrals, GTH
pseudopotentials~\cite{PhysRevB.54.1703}, and their corresponding family of
basis sets.  We use the aug-TZV2P basis set for the water molecule and a mixture
of the DZVP and TZVP basis sets for the substrate.  We use the SVZ basis as the
minimal AO basis required for projection of the occupied orbitals and the
eigenvalue cutoff is set to $0.1$ for both the occupied and virtual space,
although a number of other truncation procedures exist. We are in the process of
being exhaustively and systematically investigating optimum truncation
procedures, which may depend on the system and properties of interest; we will
publish these results elsewhere.  Following truncation, we perform post-HF
calculations using semicanonical orbitals obtained by diagonalizing the Fock
matrix in the basis of the localized fragment orbitals. 

\begin{figure}
\includegraphics[scale=0.85]{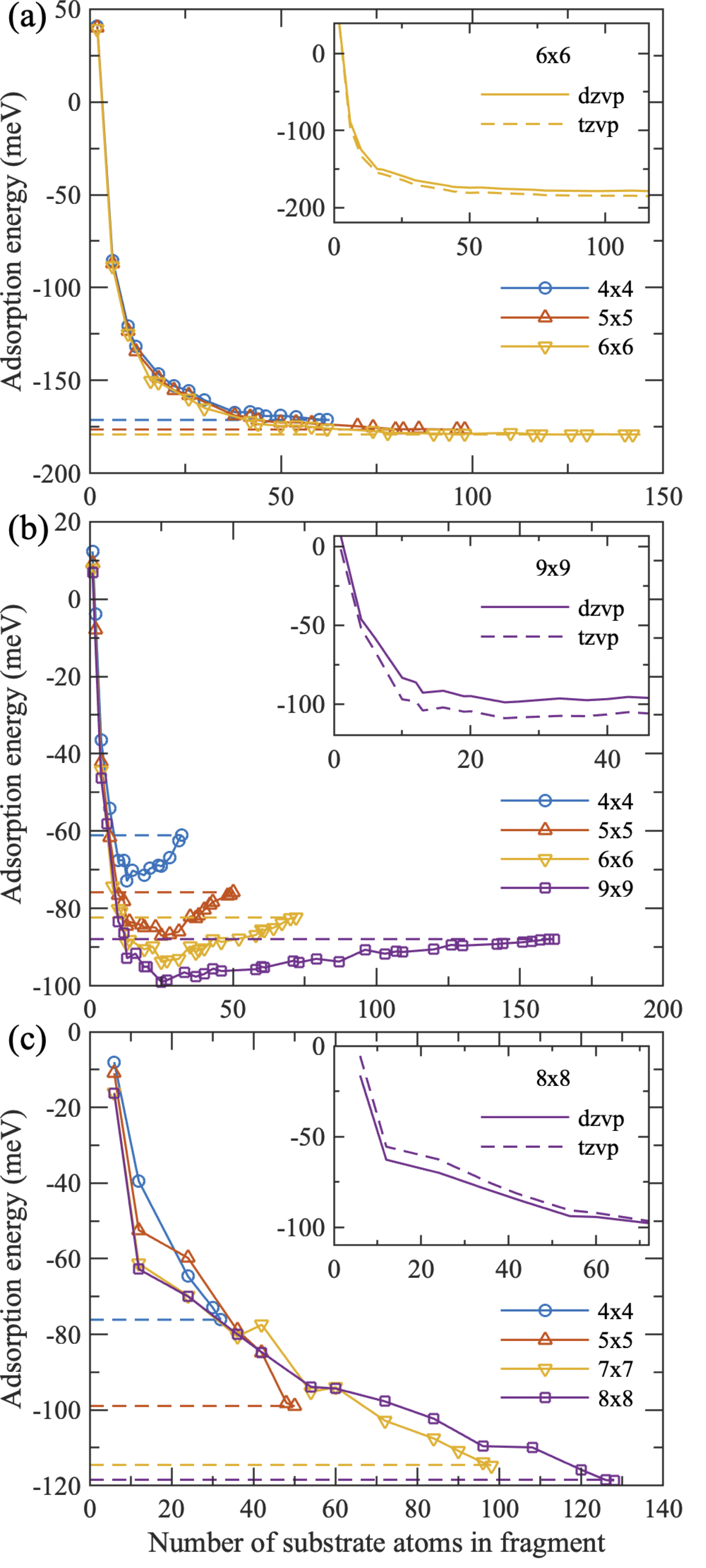}
\caption{The MP2 adsorption energy vs. fragment size, for water on (a) LiH, (b)
hBN, and (c) graphene, with the DZVP basis on the substrate. The insets compare
the DZVP and TZVP basis sets. The smaller DZVP basis allows for MP2 calculations
of all fragment sizes up to the full supercell, the results of which are
indicated by a dashed horizontal line.}
\label{fig:2}
\end{figure}

Figure~\ref{fig:2} plots the MP2 adsorption energy as a function of the number
of substrate atoms in the fragment for water on (a) LiH, (b) hBN, and (c)
graphene for a series of supercell sizes, where we use the the DZVP basis for
the substrate atoms.  The reasonable cost of MP2 calculations allows us to
increase the fragment size to the full supercell limit and thus carefully study
the convergence.  Interestingly, the convergence behaviors are quite different,
reflecting the various electronic characters of the molecule-substrate
interaction.

The adsorption energy of water on LiH, Figure~\ref{fig:2}a, rapidly converges to
the full supercell limit, and the energy minimally changes as the supercell size
is increased from $4\times 4$ to $6\times 6$. These results reflect the highly
local nature of the interaction of water with the highly insulating LiH (001)
surface, which is captured at fragment sizes much smaller than the full supercell. 

In contrast, the MP2 adsorption energy curves for hBN, Figure~\ref{fig:2}b,
quickly reach a minimum before slowly rising to the full supercell energy.  It
is possible that the minimum value represents the most accurate energy for a
given supercell size and that the rise in energy is due to error from the finite
coverage of water.  In other words, as the fragment grows in size, it includes
orbitals that are erroneously perturbed by the periodic images, leading to a
spurious interaction that raises the adsorption energy. However, as the
supercell grows in size, the water molecule is further separated from its
periodic images, and the fragment size can grow larger without including these
unphysical orbitals.  This behavior can be seen by comparing the adsorption
energy for a fixed number of atoms in the fragment, which smoothly decreases as
the supercell size is increased; importantly, the MP2 step of these calculations
have identical cost because the fragment size is the same, highlighting a
strength of our regional embedding method.

Finally, the MP2 adsorption energy of water on graphene, Figure~\ref{fig:2}c,
exhibits a very slow convergence to the the supercell limit and shows no
indication of absolute convergence.  This behavior is almost surely due to the
challenge of localizing orbitals in gapless
materials~\cite{doi:10.1021/jp972919j,RevModPhys.84.1419,PhysRevB.65.035109,Hoyvik2014}
and the closely related importance of long-range dispersion.

The inset of each panel in Figure~\ref{fig:2} shows a comparison to results
obtained by using the TZVP basis for the substrate atoms in the fragment (and
DZVP for the substrate atoms outside of the fragment).  These latter
calculations are limited in size by memory or convergence issues associated with
linear dependencies in periodic systems.  Using the larger TZVP basis on the
fragment does not qualitatively alter the convergence behavior and changes the
adsorption energy by \SI{10}{meV} or less.

\begin{figure}
\includegraphics[scale=0.85]{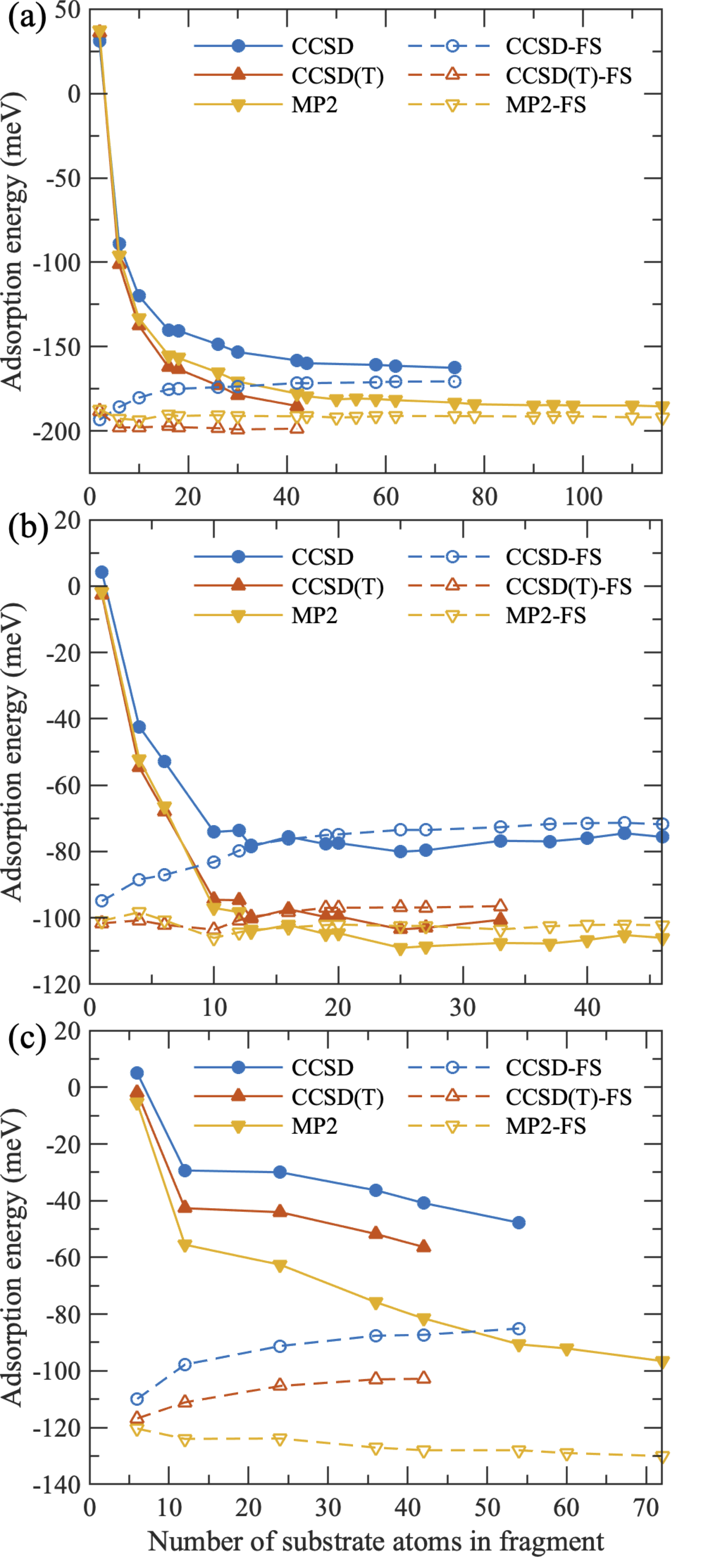}
\caption{The post-HF adsorption energy as a function of fragment size, for water
on (a) $6\times 6$ LiH, (b) $9\times 9$ hBN, and (c) $8\times 8$ graphene, with
TZVP on the substrate fragment atoms and DZVP on the rest of the substrate. The
solid lines with filled symbols correspond to uncorrected values, while the
dashed lines with open symbols correspond to finite-size (FS) corrected values,
as described in the text.
}
\label{fig:3}
\end{figure}

In Figure~\ref{fig:3}, we present our main results of the adsorption energy
obtained using MP2, CCSD, and CCSD(T), for the same three systems with the
largest supercell sizes considered, i.e.~$6\times 6$ for LiH, $9\times 9$ for
hBN, and $8\times 8$ for graphene.  As in the insets of Figure~\ref{fig:2}, all
calculations use the TZVP basis for the substrate atoms belonging to the
fragment and DZVP for the remainder, except for the (T) correction, which used
the DZVP basis for all substrate atoms,
\begin{equation}
\begin{split}
    E_\mathrm{CCSD(T)}(N,\mathrm{TZ})&=E_\mathrm{CCSD}(N,\mathrm{TZ})+\delta^{\mathrm{(T)}}(N,\mathrm{DZ}), \\
    \delta^{\mathrm{(T)}}&=\left[E_\mathrm{CCSD(T)}-E_\mathrm{CCSD}\right](N,\mathrm{DZ}),
\end{split}
\end{equation}
where $N$ is the number of fragment atoms and $E$ is the adsorption energy.
The adsorption energies, plotted with solid lines and filled symbols, follow the
same shape and rate of convergence as the MP2 results presented in
Figure~\ref{fig:2}. In particular, the shape of convergence correlates with the
adsorption energy, as seen when comparing the strong, local adsorption of water
on LiH versus the dispersion-dominated interaction with hBN and graphene.

Given the similar convergence behavior of the adsorption energies and the
relatively cheap cost of MP2, we can apply a simple finite-size (FS) correction,
\begin{equation}
\begin{split}
E(\infty,\mathrm{TZ})&=E(N,\mathrm{TZ}) + \delta^{\mathrm{FS}}(N,\mathrm{DZ}) \\
\delta^{\mathrm{FS}}(N,\mathrm{DZ}) &=  \left[E_\mathrm{MP2}(\infty)-E_\mathrm{MP2}(N)\right](\mathrm{DZ}),
\end{split}
\end{equation}
where we estimate $E_\mathrm{MP2}(\infty)$, the MP2 adsorption energy in the
thermodynamic limit by a $1/N$ extrapolation of a series of full supercell MP2
calculations (see the dashed lines in Figure~\ref{fig:2}). Although previous work has used a $1/N^2$
extrapolation of the adsorption energy, we find our data fits much better to
$1/N$ (shown in the Supporting Information).  The finite-size corrected
data are frequently converged to within 10~meV when correlating only five to ten
atoms in the substrate!
We note that the finite-size correction can also be viewed as a focal-point
basis set correction, accounting for the frozen orbitals in the supercell calculation.

\begin{table}
\begin{ruledtabular}
\begin{tabular}{ccccccc}
 & \multicolumn{2}{c}{LiH} & \multicolumn{2}{c}{hBN} & \multicolumn{2}{c}{graphene}\\
& present & lit\cite{Tsatsoulis2017}. & present & lit. & present & lit\cite{Brandenburg2019}. \\
\hline
MP2 & 192 & 233 & 102 & 110\cite{Al-Hamdani2017}, 118\cite{Gruber2018} & 130 & - \\
CCSD & 171 & 229 & 72 & 84\cite{Gruber2018} & 85 & - \\
CCSD(T) & 199 & 254 & 97 & 103\cite{Gruber2018} & 103 & 87
\end{tabular}
\end{ruledtabular}
\caption{Adsorption energy of a water molecule on LiH, hBN, and graphene, comparing
results obtained in the present manuscript and in the literature (lit.).  All energies
are in meV.}
\label{tab:ads}
\end{table}

Considering differences in implementation details (e.g.~pseudopotentials) 
and that we have not attempted a complete basis set limit extrapolation,
which can increase the adsorption energy by as much as
20~meV~\cite{Tsatsoulis2017}, our results compare favorably to recent literature
values~\cite{Tsatsoulis2017,Al-Hamdani2017,Gruber2018,Brandenburg2019}, as shown
in Table~\ref{tab:ads}.  We emphasize that previous calculations for all three
systems have only been performed on $4\times 4$ supercells and then extrapolated
to correct for finite-size effects. In contrast, our approach allows us to
directly simulate much larger supercells up to $9\times 9$, and monitor the
convergence as a function of the number of atoms in the local fragment.  Taken
together, the combination of regional embedding and focal point corrections (for
basis set and finite-size effects) enables the calculation of high-quality ab
initio adsorption energies at very low computational cost.

In summary, we have presented regional embedding, which 
allows an isolated treatment of local chemical changes with high level wave function methods.
This is achieved by 
unitary transformations of the occupied and virtual Hartree-Fock orbitals based on the
degree of overlap with the fragment's Hilbert space.
The conventional use of high level wave functions in sufficiently large supercells is expensive.
In contrast, regional embedding restricts the correlated calculation to the
Hilbert space associated with the target fragment; consequently, large supercells can be used 
and the finite-size error can be efficiently eliminated.
The benchmarks presented indicate that regional embedding with finite-size (focal point) corrections allows 
remarkably quick convergence to the thermodynamic limit, although the rate of convergence depends on the
nature of molecule-substrate interaction.  Converged results are 
achieved with correlated calculations on much smaller systems than the supercells used in contemporary 
periodic quantum chemistry and quantum Monte Carlo methods.

We expect this method to have broad applicability in molecular, biological, and
condensed systems.  The simplicity and performance should make it an appealing
choice whenever modeling localized chemical changes in large systems.  This
covers \emph{many} applications scenarios, and we have already started projects
in heterogeneous catalysis and in the optical and thermochemical properties
crystal defects.  Concerning the theory, treating the frozen orbitals at levels
beyond focal-point corrections may further increase the power of the method;
self-consistent quantum embeddings, canonical transformations, or screened
interactions may provide viable avenues.  Finally, a core appeal of this method
is that the embedded quantum system can be handled with essentially arbitrary
many-body solvers, including quantum chemical multireference methods for
strongly correlated systems.  This may open a pathway to reliable treatments of
complex electronic structure phenomena also in condensed phases and at
interfaces.  For example, these methods can be used to address long-standing
problems in catalyst doping, catalysis at surface defects (e.g.~kinks and
edges), or the surface chemistry of transition metals and radicals.

\section*{Associated Content}
\subsection*{Supporting Information}
The Supporting Information is available free of charge at [xxx].
Finite-size extrapolation of MP2 adsorption energies and discussion of the choice
of eigenvalue cutoff in the regional embedding projection operators.

\section*{Acknowledgments}
T.C.B.~thanks Sandeep Sharma for helpful discussions.  This work was supported
in part by the NSF CAREER program under award No.~CHE-1945276 (G.K.) and by the
NSF Cyberinfrastructure for Sustained Scientific Innovation program via award No.~OAC-1931258 (T.C.B.).
The Flatiron Institute is a division of the Simons Foundation.

\section*{References}
\bibliographystyle{achemso}
\bibliography{manuscript_bib_1,manuscript_bib_2}

\end{document}


\title{Supporting Information: Regional Embedding Enables High-Level Quantum Chemistry for Surface Science}

\author{Bryan T. G. Lau}
\affiliation{Center for Computational Quantum Physics, Flatiron Institute, New York, New York 10010 USA}
\author{Gerald Knizia}
\email{knizia@psu.edu}
\affiliation{Department of Chemistry, Pennsylvania State University, University Park, Pennsylvania 16802 USA}
\author{Timothy C. Berkelbach}
\email{tim.berkelbach@gmail.com}
\affiliation{Center for Computational Quantum Physics, Flatiron Institute, New York, New York 10010 USA}
\affiliation{Department of Chemistry, Columbia University, New York, New York 10027 USA}

\maketitle

\renewcommand\theequation{S\arabic{equation}}

\onecolumngrid
\section*{Extrapolation of MP2 adsorption energy curves}

\begin{figure}[h]
\includegraphics[scale=0.8]{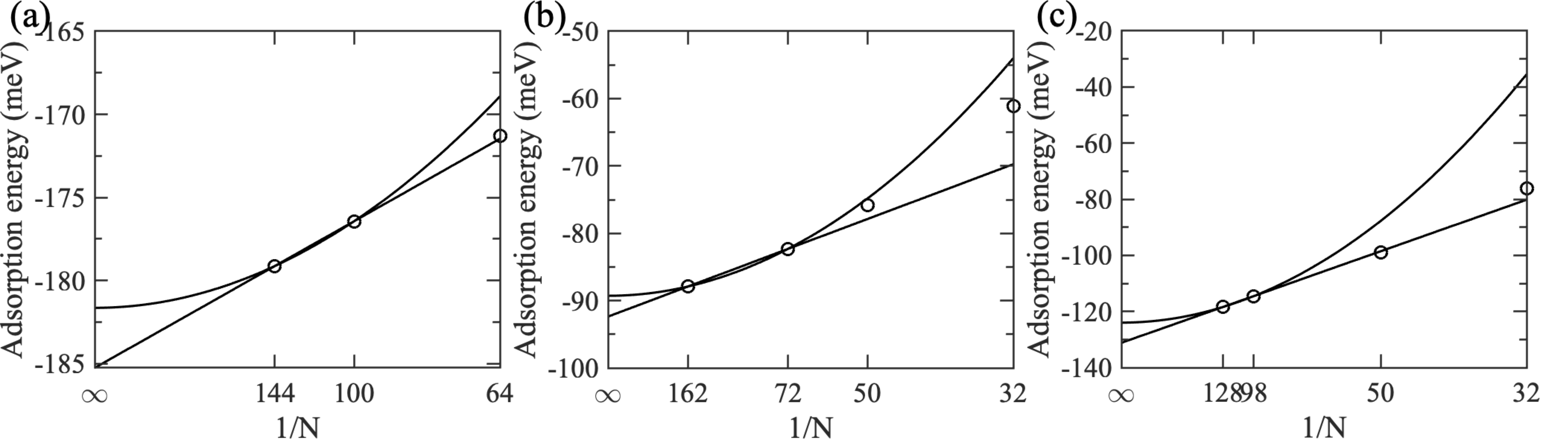}
\caption{The supercell MP2 adsorption energies for (a) LiH, (b) hBN, and (c)
graphene. The energies are fit to a linear and quadratic form, where $N$ is the
number of substrate atoms in the supercell.
}
\label{fig:SupFig1}
\end{figure}

Figure~\ref{fig:SupFig1} plots the supercell MP2 adsorption energies for
the systems studied in the manuscript, using the DZVP basis on the substrate. 
We compare fits of the energies to a linear $1/N$ and
quadratic $1/N^2$ form to extrapolate the results to the thermodynamic, low-density limit.
We see that $1/N$ extrapolation fits our data best, although different extrapolations
produce very similar adsorption energies, differing by only 5-8~meV.

